\documentclass[aps,prl,preprint,preprintnumbers,amsmath,amssymb]{revtex4-1}
\usepackage{graphicx}
\usepackage{dcolumn}
\usepackage{bm}
\usepackage{amsmath}
\usepackage{color}

\begin{document}

\def\gsim{\mathop {\vtop {\ialign {##\crcr 
$\hfil \displaystyle {>}\hfil $\crcr \noalign {\kern1pt \nointerlineskip } 
$\,\sim$ \crcr \noalign {\kern1pt}}}}\limits}
\def\lsim{\mathop {\vtop {\ialign {##\crcr 
$\hfil \displaystyle {<}\hfil $\crcr \noalign {\kern1pt \nointerlineskip } 
$\,\,\sim$ \crcr \noalign {\kern1pt}}}}\limits}


\title{
%
Topological magnetic textures and long-range orders in Tb-based quasicrystal and approximant
}


\author{Shinji Watanabe}
\affiliation{Department of Basic Sciences, Kyushu Institute of Technology, Kitakyushu, Fukuoka 804-8550, Japan 
}


\date{\today}

\begin{abstract} 
The quasicrystal(QC)s have unique lattice structure with the rotational symmetry forbidden in the periodic crystals. The electric properties are far from complete understanding. It has been unresolved whether the magnetic long-range orders are realized in the QC. Here we report our theoretical discovery of the ferromagnetic long-range order in the Tb-based QC. The difficulty in past theoretical studies on the QC was lack of the microscopic theory of the crystalline electric field (CEF), which is crucially important in the rare-earth systems. By analyzing the CEF in the Tb-based QC, we clarify that magnetic anisotropy plays a key role in realizing unique magnetic textures in the Tb-based QC and approximant crystal (AC). By constructing the minimal model, we show that various magnetic textures on the icosahedron at whose vertices Tb atoms are located. We find that the hedgehog state is characterized by the topological charge of one and the whirling-moment state is characterized by unusually large topological charge of three. The hedgehog and whirling-moment states are shown to be realized as antiferromagnetic orders transcribed as the emergent monopole and antimonopole in the 1/1 AC. We find that these states exhibit the topological Hall effect under applied magnetic field accompanied by the topological as well as metamagnetic transition. Our model and the determined phase diagram are expected to be relevant to the broad range of the rare-earth based QCs and ACs with strong magnetic anisotropy, which are useful not only to understand magnetism but also to explore novel topological properties. 
\end{abstract}


\maketitle

{T}he quasicrystal(QC)s have unique lattice structure with rotational symmetry forbidden in periodic crystals~\cite{Shechtman,Tsai,Takakura}. 
In the QC, it has been unresolved whether the magnetic long-range order is realized~\cite{Suzuki}. 
In the 1/1 AC composed of rare earths, which is the periodic crystal with the same local atomic configuration with the QC, 
magnetic long-range orders have been observed in 
Cd$_6$R (R=Pr, Nd, Sm, Gd, Tb, Dy, Ho, Eu, and Tm)~\cite{Tamura2010,Mori,Tamura2012} and Au-SM-R (SM=Si, Al, Ge, and Sn; R=Gd, Tb, Dy, and Ho)~\cite{Hiroto2013,Hiroto2014,Das}. 
Among them, the magnetic structures have recently been identified experimentally as 
the ferromagnetic (FM) order in Au$_{70}$Si$_{17}$Tb$_{13}$~\cite{Hiroto} and antiferromagnetic (AFM) order in Au$_{72}$Al$_{14}$Tb$_{14}$~\cite{Sato2019}. 
However, the ordering mechanism remains elusive although theoretical analyses based on the spin model and the Hubbard model in mostly small clusters or low spatial-dimension systems were reported~\cite{Axe,Wessel,Kons,Jan2007,Hucht,Thiem,Komura,Sugimoto,Koga2017,STS}, 
where the magnetic anisotropy originating from the crystalline electric field (CEF) at the rare earth was not taken into account microscopically. 
Moreover, recent experimental discovery of the FM long-range order in the Tb-based QC~\cite{Tamura2021} has opened a new stage of research, which also calls for theoretical study from this viewpoint. 

In this report, we present our theoretical discovery of topological magnetic textures and magnetic long-range orders in the Tb-based 1/1 AC and QC. 
We clarify that the valences of the ligand ions surrounding Tb play a key role in controlling the magnetic anisotropy in the CEF, which is crucial to drive unique magnetic textures in the AC and QC. 
We find the hedgehog state characterized by topological charge $n=1$ and 
also the whirling-moment states characterized by unusually large topological charge $n=3$. These topological states 
are shown to be realized as the AFM orders with emergent monopole and antimonopole in the 1/1 AC, which exhibit topological Hall effect under applied magnetic field accompanied by the topological as well as metamagnetic transition.

\section*{Results}

\subsection*{Crystalline electric field}
\begin{figure}[t]
\includegraphics[width=7.5cm]{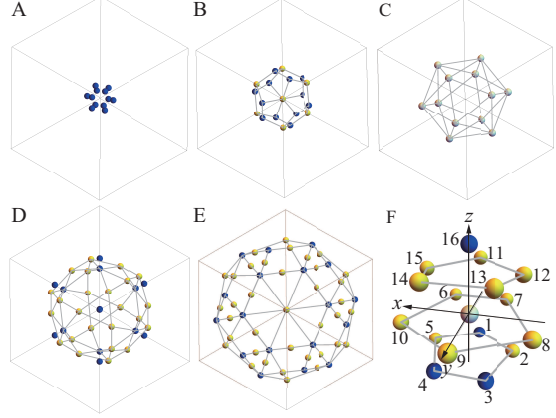}%
\caption{(color online) 
Tsai-type cluster consists of (A) cluster center, (B) dodecahedron, (C) IC, (D) icosidodecahedron, and (E) defect rhombic triacontahedron 
with Tb (gray), Si (blue), and Au (yellow). 
(F) Local configuration around Tb. 
The number labels surrounding Si and Au. 
}
\label{fig:atoms}
\end{figure}

Let us start with the CEF analysis of the Tb-based QC and AC by considering the Au-SM-R system. 
The 1/1 AC Au-SM-R consists of the Tsai-type cluster with concentric shell structures as illustrated in Figs.~\ref{fig:atoms}A-\ref{fig:atoms}E where Au$_{70}$Si$_{17}$Tb$_{13}$ is shown as a typical case~\cite{Hiroto}.
In Fig.~\ref{fig:atoms}C, Tb atoms are located at each vertex of the icosahedron (IC), forming the Tb 12 cluster. 
The local configuration of atoms surrounding Tb is shown in Fig.~\ref{fig:atoms}F where the $z$ axis is taken as the direction passing through a Tb site from the center of the IC and the $y$ axis is taken so as the $yz$ plane to be the mirror plane. 

Since Tb$^{3+}$ has $4f^{8}$ configuration i.e., more than half of the closed 4f shell $(4f^{14})$, 
the CEF Hamiltonian for Tb$^{3+}$ is expressed as $H_{\rm CEF}=6|e|V_{\rm cry}$ in the hole picture on the basis of the point charge model. Here, the potential $V_{\rm cry}$ is given by
%
$
V_{\rm cry}=\sum_{i=1}^{16}q_i/|{\bm R}_i-{\bm r}|, 
$
%
where ${\bm R}_i$ is the position vector of the ligand ions Au$^{{\rm Z}_{\rm Au}+}$ and Si$^{{\rm Z}_{\rm Si}+}$ with $Z_{\rm Au}$ and $Z_{\rm Si}$ being the valences of Au and Si respectively. 
Then the charge $q_i$ is given by $q_i=Z_{\rm Au}|e|$ and $q_i=Z_{\rm Si}|e|$ on the $i$th site in Fig.~\ref{fig:atoms}F.

Recently, $H_{\rm CEF}$ in rare-earth based QCs and ACs has been formulated by the operator equivalents~\cite{WM2021} as 
\begin{eqnarray}
H_{\rm CEF}=
\sum_{\ell=2,4,6}\left[B_{\ell}^{0}(c)O_{\ell}^{0}(c)+\sum_{\eta=c,s}\sum_{m=1}^{\ell}B_{\ell}^{m}(\eta)O_{\ell}^{m}(\eta)\right]. 
\label{eq:H}
\end{eqnarray}
Here, $O_{\ell}^{m}(c)$ and $O_{\ell}^{m}(s)$ are the Stevens operators~\cite{Stevens} and $B_{\ell}^{m}(\eta)$ is  the coefficient. 

\begin{figure}[t]
\includegraphics[width=6.5cm]{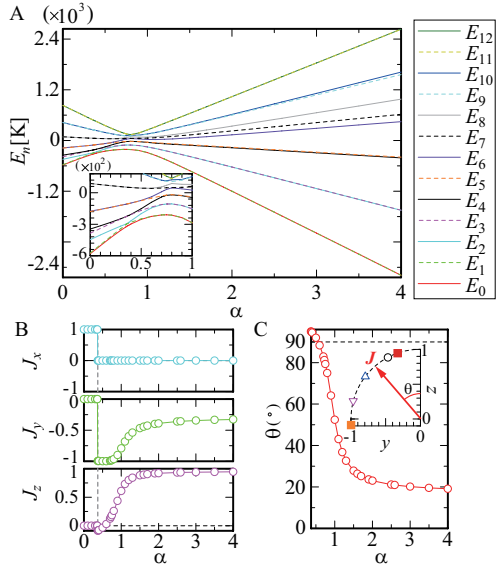}%
\caption{(color online) 
(A) The $\alpha$ dependence of the CEF energy lebels $E_n$ for $Z_{\rm Si}=\alpha Z_{\rm Au}$ with $Z_{\rm Au}=0.223$. 
Inset shows the enlargement for $0\le \alpha\le 1$. 
(B) The $\alpha$ dependence of the $x$ (top panel), $y$ (middle panel), and $z$ (bottom panel) components of the largest magnetic moment for the CEF ground state. 
(C) The $\alpha$ dependence of the momentum direction $\theta$ defined as angle from the $z$ axis in the $y$--$z$ plane (see inset) for $0.37\le \alpha\le 4.0$. In inset, the points directed by ${\bm J}$ are indicated for $\alpha=0.37$ (orange square), $\alpha=0.8$ (purple inverted triangle), $\alpha=1.0$ (blue triangle), $\alpha=1.5$ (black circle), and $\alpha=4.0$ (red square). 
}
\label{fig:E_CEF}
\end{figure}

In metallic crystals, the valences of Au and Si are known to be 1 and 4 respectively in normal metals~\cite{Pearson} and alloyed AC~\cite{Mizutani}. In reality, screening effect by conduction electrons reduces $Z_{\rm Au}$ and $Z_{\rm Si}$ from these values. 
Hence, we analyze the CEF energies for $0\le \alpha\le 4$ in $Z_{\rm Si}=\alpha Z_{\rm Au}$. Here, $Z_{\rm Au}$ is set to be 0.223 as a typical value, 
which was determined by the neutron measurement in Au$_{70}$Si$_{17}$Tb$_{13}$~\cite{Hiroto}. 
By diagonalizing $H_{\rm CEF}$ for the total angular momentum $J=6$ as the ground multiplet by the Hund's rule, we obtain the CEF energies $E_n$ for $n=0$--$12$ as shown in Fig.~\ref{fig:E_CEF}A. 
Note here that the choice of $Z_{\rm Au}$ itself does not affect the eigenstates of $H_{\rm CEF}$ $|\psi_n\rangle$ as far as $\alpha$ is the same.
The CEF energies are seen to split into almost 7 to 8 levels, some of which are nearly degenerate. 
The energy difference between the ground state and the first-excited state is typically in the order of $10^2$~K to $10^3$~K. Hence, the CEF ground state is dominant over the lower-temperature properties.

Next, to clarify the principal axis of the magnetic moment in the CEF ground state, we calculate $3\times 3$ matrix $M_{\xi,\zeta}\equiv \langle\psi_0|\hat{J}_{\xi}\hat{J}_{\zeta}|\psi_0\rangle$ for $\xi, \zeta=x, y,$ and $z$, where $\hat{J}_{\xi}$ is the operator of the total angular momentum. 
By diagonalizing $M$, we obtain the normalized eigenvector for the largest eigenvalue, which gives the largest moment direction ${\bm J}=(J_x, J_y, J_z)$. 
We find that the moment direction changes depending on $\alpha$ as shown in Fig.~\ref{fig:E_CEF}B.
For $0.37\le \alpha\le 4.0$, the moment is lying in the mirror plane i.e. the $yz$ plane in Fig.~\ref{fig:atoms}F, 
while for $0\le \alpha<0.37$, the moment is directed as ${\bm J}\parallel (1,0,0)$. 
At $\alpha=0.37$, ${\bm J}$ points to the $\theta=95.06^{\circ}$ direction from the $z$ axis
and as $\alpha$ increases $\bm J$ rotates to anticlock-wise direction and tends to approach $\theta=0^{\circ}$ i.e., the pseudo 5-fold axis [$z$ axis in Fig.~\ref{fig:atoms}F]. 

\subsection*{Minimal model with magnetic anisotropy}

To clarify how magnetic anisotropy by the CEF affects magnetism on the IC, we consider the minimal model
\begin{eqnarray}
H=-\sum_{\langle i,j\rangle}J_{i,j}\hat{\bm J}_i\cdot\hat{\bm J}_j,
\label{eq:HI}
\end{eqnarray}
where $J_{ij}$ is exchange interaction between the $i$th and $j$th Tb site on the IC [see Fig.1(c)]. Here, $\hat{\bm J}_i$ represents the unit Ising-spin-vector operators of Tb$^{3+}$ whose direction is restricted to either parallel or antiparallel to the moment direction shown in Fig.2B. 
We consider the nearest neighbor (N.N.) interaction $J_1$ 
and next N.N. (N.N.N.) interaction $J_2$. 
So far, in several Tb-based ACs, positive Weiss temperature has been observed, which indicates the FM interaction between the magnetic moments at Tb sites~\cite{Suzuki}. 
Actually, in the Tb-base QC where the FM long-range order has been observed, the positive Weiss temperature is observed~\cite{Tamura2021}. 
Hence, we focus on the FM interactions $J_1>0$ and $J_2>0$. 
The case for the AFM interaction will be reported elsewhere.

\begin{figure}
\centering
\includegraphics[width=14cm]{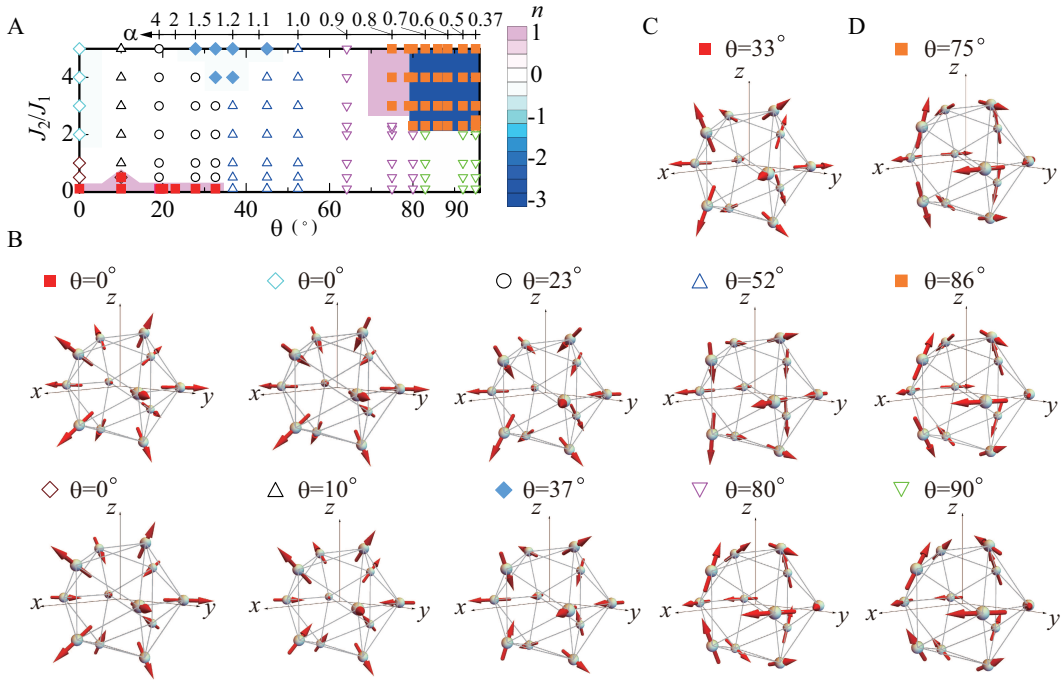}%
\caption{(color online) 
(A) The ground-state phase diagram of the model (\ref{eq:HI}) on an IC and the 1/1 AC for $J_2/J_1$ with $J_1>0$ and $\theta$ as well as $\alpha$. 
In the 1/1 AC, open (filled) symbols indicate the FM (AFM) orders of the magnetic IC. 
Contour plot of the topological charge $n$ on an IC is also shown. 
(B) Magnetic textures of each symbol in (A). Hedgehog state (red square) and 
antiwhirling state (orange square). 
(C) Hedgehog state for $\theta=33^{\circ}$. 
(D) Antiwhirling state for $\theta=75^{\circ}$ characterized by $n=1$. 
}
\label{fig:PD}
\end{figure}

By performing numerical calculations, we determine the ground-state phase diagram in Fig.~\ref{fig:PD}A. 
Here we plot the wide range of $\theta$ as well as $\alpha$ for the horizontal axis. If we consider the effects of the Au-SM mixed sites in the Au-SM-Tb QCs and ACs, the $\alpha$ dependence of $\theta$ is expected to change slightly from that in Fig.~\ref{fig:E_CEF}C~\cite{WM2021}. Hence, the $J_2/J_1$--$\theta$ phase diagram is relevant generally to the rare-earth based QC and AC with strong magnetic anisotropy. 
The phase diagram for $0\le\alpha\le 0.37$ (see Fig.~\ref{fig:E_CEF}B) is shown in SI (Fig.~S1). 

\subsection*{Magnetic textures on icosahedron and topological charges}

In Fig.~\ref{fig:PD}A, 
we find that unique magnetic textures appear depending on $J_2/J_1$ and $\theta$ as denoted by each symbol, which is shown in Fig.~\ref{fig:PD}B. 
We find that the hedgehog state~\cite{STS} (red square $\theta=0^{\circ}$ in Fig.~\ref{fig:PD}B), where all the moments at the 12 Tb sites on the IC are directed outword, 
appears in the wide range of $\theta$ for $0\le\theta\le 33^{\circ}$ for $J_2/J_1=1/10$ in Fig.~\ref{fig:PD}A. 
The hedgehog state for $\theta=33^{\circ}$ is shown in Fig.~\ref{fig:PD}C. 
In the hedgehog states, the total magnetic moment on the IC ${\bm J}_{\rm tot}=\sum_{i=1}^{12}\langle\hat{\bm J}_i\rangle$ is zero. 
The ${\bm J}_{\rm tot}={\bf 0}$ state is also realized 
as the whirling-moment state denoted by orange squares in Figs.~\ref{fig:PD}A and \ref{fig:PD}B where the magnetic moments are whirling if they are seen from the (111) direction (SI, Fig.~S2). 


To clarify the topological character of each magnetic texture, 
we define 
the scalar chirality of the IC with its center-of-mass position ${\bm R}$ by $\chi({\bm R})=\sum_{i,j,k\in{\rm IC}}\chi_{i,j,k}$ with $\chi_{i,j,k}\equiv {\bm J}_i\cdot({\bm J}_j\times{\bm J}_k)$ where the order of $i$, $j$, and $k$ is defined in the anticlockwise direction with respect to the normal vector of the triangle formed by the $i$, $j$, and $k$th sites, $\hat{n}_{i,j,k}$, pointing outward from ${\bm R}$. 
In the same way, we define the solid angle subtended by the twelve moments on the IC as 
$\Omega({\bm R})=\sum_{i,j,k\in{\rm IC}}\Omega_{ijk}$. 
Here, $\Omega_{ijk}$ is the solid angle for three magnetic moments ${\bm J}_i$, ${\bm J}_j$, and ${\bm J}_k$, 
which is given by $\Omega_{ijk}=2\tan^{-1}[\chi_{ijk}/(1+{\bm J}_i\cdot{\bm J}_j+{\bm J}_j\cdot{\bm J}_k+{\bm J}_k\cdot{\bm J}_i)]$~\cite{Eriksson}. 
The topological charge $n$ is defined as $n\equiv\Omega({\bm R})/(4\pi)$ per an IC. 
Then we find that the hedgehog is characterized by $n=1$. This implies that the magnetic moments at the 12 Tb sites on the IC cover all the sphere once, which is regarded as the emergent monopole acting as the source of emergent field~\cite{Dirac,Volovik,Kotiuga,Feldtkeller,Doring,Kabanov,Thiaville,Milde,Castelnovo,Morris,Aoyama}. 
Interestingly, we find that unusually large magnitude of the topological charge $n=-3$ is realized in the 
whirling state in the large-$\theta$ region for $\theta\gsim 83^{\circ}$ and $J_2/J_1>2$ (orange square) (we call this state the 
antiwhirling state hereafter) in Fig.~\ref{fig:PD}A. 
Furthermore, we find that the topological transition from $n=-3$ to $n=1$ occurs around $\theta=79.5^{\circ}$ (see Fig.~\ref{fig:PD}D) in the antiwhirling region in Fig.~\ref{fig:PD}A. 

We also show the contour plot of the topological charge $n$ in Fig.~\ref{fig:PD}A. 
The finite topological charge with integer $n\ne 0$ appears in the ${\bm J}_{\rm tot}={\bf 0}$ states. 
As distinct from topological spin textures intensively studied in periodic crystals~\cite{Nagaosa,Kanazawa,Tokura,Aoyama}, 
our finding is that the total angular momentum, i.e., the orbital angular momentum coupled to spin, forms the novel topological textures  designed by the CEF-anisotropy protected IC.


\subsection*{Magnetism in 1/1 approximant}

Next, let us consider the Tb-based 1/1 AC with the body-centered-cubic (bcc) lattice structure composed of the Tsai-type cluster. 
In the unit cell, there are two ICs, one of which is located at the bcc center and the other is at the bcc corner. 
When the model (\ref{eq:HI}) is applied to the 1/1 AC, where $\langle i,j\rangle$ in Eq.~(\ref{eq:HI}) is not only taken into account for the intra-IC Tb pairs but also for the inter-IC Tb pairs, we find that the N.N. (N.N.N.) inter-IC $\langle\hat{\bm J}_i\cdot\hat{\bm J}_j\rangle$ equals to the N.N.N. (N.N.) intra-IC $\langle\hat{\bm J}_i\cdot\hat{\bm J}_j\rangle$. 
In the AC Au$_{70}$Si$_{17}$Tb$_{13}$ as a typical case, the nearest-neighbor (N.N.) Tb distance in the intra IC is $0.38a$ and the next N.N. (N.N.N.) Tb distance is $0.61a$ where $a=14.726~{\rm \AA}$ is the lattice constant of the bcc unit cell~\cite{Hiroto}. On the other hand, the N.N. and N.N.N. Tb distances for the inter ICs are $0.37a$ and $0.53a$ respectively. Hence, the N.N. Tb distances for the intra IC and inter IC are close. As for the N.N.N. Tb distances, both values for the intra IC and inter IC are also close. 
Hence, if the inter-IC N.N. and N.N.N. interactions are set to be $J_1$ and $J_2$ respectively as done for the intra-IC N.N. and N.N.N. interactions respectively, 
either of the FM or AFM arrangement of the magnetic IC 
at the bcc center and corner can be determined as the ground state by evaluating the energy of the inter-IC contributions in Eq. (\ref{eq:HI}). 

\begin{figure}[t]
\includegraphics[width=8.5cm]{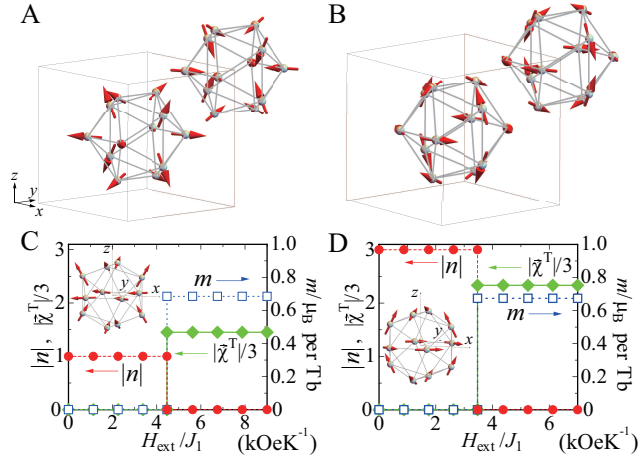}%
\caption{(color online) 
In the 1/1 AC, at the bcc center and corner in the unit cell, 
(A) hedgehog state ($n$=1) and antihedgehog state ($n$=$-1$)  for $\theta=28^{\circ}$ 
and 
(B) antiwhirling state ($n$=$-3$) and whirling state ($n$=3) for $\theta=86^{\circ}$ 
are located respectively. 
(C), (D) The magnetic-field dependence of the topological charge $|n|$, total chirality $|{\bm \chi}^{\rm T}|$, and magnetization $m$ applied to (A), (B) respectively with $J_1$ being the N.N. interaction in unit of K. Insets in (C) and (D) show Tb moments on the IC for $H_{\rm ext}>H_{\rm M}$. 
}
\label{fig:Tb_AC}
\end{figure}

The result for the 1/1 AC is shown in Fig.~\ref{fig:PD}A, where the FM (AFM) order of the magnetic IC is denoted by the open (filled) symbols. 
The ${\bm J}_{\rm tot}={\bm 0}$ states 
are realized as all AFM orders. 
Namely, the hedgehog $(n=1)$ state and antihedgehog $(n=-1)$ state where all the moments are inverted from the hedgehog are located at the bcc center and corner respectively as shown in Fig.~\ref{fig:Tb_AC}A. 
The whirling state $(n=3)$ where all the moments are inverted from the antiwhirling state and the antiwhirling state $(n=-3)$ at the bcc corner and center are realized respectively as in Fig.~\ref{fig:Tb_AC}B. 
These are regarded as emergent monopole and antimonopole with the ``charge'' $n$, acting as the source and sink of emergent field respectively. 

It is noteworthy that neutron measurements with the analysis based on the model (\ref{eq:HI}) on an IC identified $\theta=86^{\circ}$ and $J_2/J_1=2.3$ for Au$_{72}$Al$_{14}$Tb$_{14}$ showing the AFM order (orange square in Fig.~\ref{fig:PD}B)~\cite{Sato2019} and $\theta=80^{\circ}$ and $J_2/J_1>0$ for Au$_{70}$Si$_{17}$Tb$_{13}$ showing the FM order (pink inverted triangle in Fig.~\ref{fig:PD}B)~\cite{Hiroto}, which are consistent with our results in Fig.~\ref{fig:PD}A. 
This indicates that the magnetic anisotropy arising from the CEF i.e. $\theta$ plays a key role in stabilizing the AFM or FM order in the 1/1 AC. 
Furthermore, the 
former 
is revealed to be characterized by the unusually large topological charge $|n|=3$. 
These results indicate that by slightly changing the ratio of valences of ligand ions $\alpha=Z_{\rm Si(Al)}/Z_{\rm Au}$, 
the magnetic and topological states can be switched. This 
is feasible by controlling the compositions of rare-earth based AC and QC.

\subsection*{Topological Hall effect}

Notable is that intriguing phenomena such as the topological Hall effect are expected to emerge in the magnetic textures in Fig.~\ref{fig:PD}A. 
The topological Hall conductivity $\sigma_{\mu\nu}^{\rm T}$ is proportional to the total magnetic chirality multiplied by a geometrical factor 
${\bm \chi}^{\rm T}=\sum_{\langle i,j,k\rangle}\chi_{ijk}\hat{n}_{ijk}$, i.e., $\sigma_{\mu\nu}^{\rm T}\propto\epsilon_{\mu\nu\rho}\chi_{\rho}^{\rm T}$~\cite{Tatara}, where $\sum_{\langle ijk\rangle}$ denotes the summation over all the three sites on each IC and $\hat{n}_{ijk}$ represents the surface normal. Here, ${\bm \chi}^{\rm T}$ plays a role as 
emergent fictitious magnetic field. 
In the hedgehog state (red square) and antiwhirling state (orange square) in Fig.~\ref{fig:PD}A, 
we confirmed ${\bm \chi}^{\rm T}={\bm 0}$.

Then, we apply magnetic field ${\bm H}_{\rm ext}$ to the 1/1 AC as 
${\cal H}=H-g_{J}\mu_{\rm B}\sum_{i}\hat{\bm J}_i\cdot{\bm H}_{\rm ext}$, where $g_{J}$ is Land{\'e}'s $g$ factor and $\mu_{\rm B}$ is the Bohr magneton. 
The result for ${\bm H}_{\rm ext}\parallel (0,0,1)$ applied to the hedgehog-antihedgehog AF ordered state (Fig.~\ref{fig:Tb_AC}A) is shown in Fig.~\ref{fig:Tb_AC}C. 
We find that the metamagnetic transition occurs at $H_{\rm M}/J_1=3.47$~kOeK$^{-1}$, where the moments at half of Tb sites in an IC are flipped, 
giving rize to the change in the topological charge $|n|$=1~$\to$~0 as well as ${\bm \chi}^{\rm T}={\bm 0}$~$\to$~$(-4.35,0,0)$. 
This implies the topological Hall effect to be observed in $\sigma_{yz}^{\rm T}$ for $H_{\rm ext}>H_{\rm M}$. 

The result for the whirling-antiwhirling AF ordered state is shown in Fig.~\ref{fig:Tb_AC}D. 
At $H_{\rm M}/J_1=4.46$~kOeK$^{-1}$, metamagnetic transition occurs as observed in Au$_{72}$Al$_{14}$Tb$_{14}$~\cite{Sato2019}. 
We find that the change in the topological charge $|n|$=3~$\to 0$ as well as the total chirality ${\bm \chi}^{\rm T}$=${\bm 0}$~$\to(-6.32,0,3.01)$ occurs at $H_{\rm M}$. This implies the topological Hall effect to be observed in $\sigma_{yz}^{\rm T}$ and $\sigma_{xy}^{\rm T}$ for $H_{\rm ext}>H_{\rm M}$.

\subsection*{Magnetism in quasicrystal}

To explore the magnetic long-range order in the QC, 
we apply the model (\ref{eq:HI}) to the Cd$_{5.7}$Yb-type QC~\cite{Takakura}, where the Tb-12 cluster, i.e., IC, is located at the origin surrounded by 30 ICs located at each vertex of the $\tau^3$-times enlarged icosidodecahedron from that in Fig.~\ref{fig:atoms}D and they are repeatedly arranged outward in the self-similar manner. Here $\tau=(1+\sqrt{5})/2$ is the golden ratio. 
In the model~(\ref{eq:HI}) applied to the QC, 
we set the FM N.N. interaction $J_1>0$ and N.N.N. interaction $J_2>0$ not only for the intra IC but also for the inter ICs which are the neighboring pairs on the line connected at each vertex of the icosidodecahedron (see Fig.~\ref{fig:atoms}D). 
Then we find that the FM long-range order is realized for $64.1^{\circ}\lsim\theta\lsim 80^{\circ}$ (at least in the region shown in Fig.~\ref{fig:QC_mom}A), as shown in Fig.~\ref{fig:QC_mom}B. 
The total magnetic moment in the IC ${\bm J}_{\rm tot}$ points to the (111) direction. 

It is noted that by applying the model (2) with the AFM interactions $J_1<0$ and $J_2<0$ to the Cd$_{5.7}$Yb-type QC, we have confirmed that there exists the region of $\theta$-$J_2/J_1$ in the ground-state phase diagram where the uniform arrangement of the hedgehog state is stabilized.

\section*{Discussion}

\begin{figure}[t]
\includegraphics[width=6.5cm]{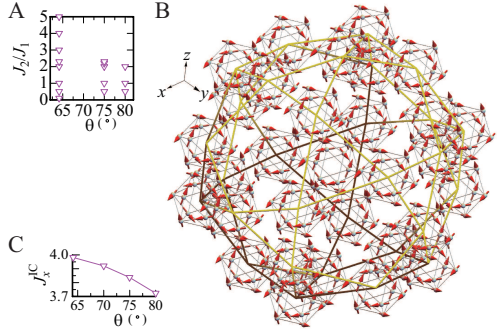}%
\caption{(color online) 
(A) FM long range order in QC realized in the $\theta$-$J_2/J_1$ plane. 
(B) The magnetic moments in the Tb-12 clusters at the origin and at the vertices of the icosidodecahedron are shown for $\theta=80^{\circ}$. 
Green (brown) lines at front (back) side connects the vertices of the icosidodecahedron. 
(C) The $\theta$ dependence of the total magnetic moment in the IC ${\bm J}_{\rm tot}=(J_x^{\rm IC},J_x^{\rm IC},J_x^{\rm IC})$ in the FM phase. 
}
\label{fig:QC_mom}
\end{figure}

We have discovered that the magnetic state on the IC denoted by pink inverted triangle in Fig.~\ref{fig:PD}B orders ferromagnetically in the QC, as shown in Fig.~\ref{fig:QC_mom}B. 
On the other hand, the hedgehog (whirling) state encounters huge degeneracy in the ground state because of strong frustration to realize the hedgehog-antihedgehog (whirling-antiwhirling) AF order due to the triangular network in the icosidodecahedron (see Fig.~\ref{fig:QC_mom}B and also Fig.~\ref{fig:atoms}D). 
Namely, within the model (2), the AFM states are degenerated. This is reminiscent of the classical spin 1/2 Heisenberg model on the triangular lattice where the macroscopic degeneracy remains in the ground state~\cite{Wannier}. 
This is in sharp contrast to the AFM order realized in the 1/1 AC with the N.N. FM interaction which forms the bipartite lattice in terms of the IC. 
To lift the degeneracy, it is necessary to take into account the effect beyond the model (2) such as transverse components of the exchange interaction and/or the longer-range interactions as the RKKY interaction. Such an analysis will be important for each material by considering the material dependent factors, which is left for the next step of the study. 
Melting the hedgehog-antihedgehog (whirling-antiwhirling) AF order by 
considering these effects under 
the frustration in the QC, which may cause non-trivial topological liquid analog to the quantum spin liquid, calls for an interesting future study.

\section*{Materials and Methods}

\subsection*{Quasicrystal and approximant crystal}

The QC and the AC consist of the Tsai-type cluster shown in Figs.~\ref{fig:atoms}A-\ref{fig:atoms}E. 
The AC retains the periodicity as well as the local atomic configuration common to the QC. 
There exist a series of ACs such as 1/1 AC, 2/1 AC, 3/2 AC, $\cdots$, where the $n\to\infty$ limit in the $F_{n-1}/F_{n-2}$ AC corresponds to the QC with $F_{n}$ being the Fibonacci number. 
In the rare-earth based 1/1 AC composed of the Tsai-type cluster, there exist two ICs in the unit cell of the body-center-cubic (bcc) lattice, where the rare-earth atoms are located at the twelve vertices of each IC. 

\subsection*{Analysis of the crystalline electric field}

In the CEF Hamiltonian~(\ref{eq:H}), 
the coefficient $ B_{\ell}^{m}(\eta)$ is given by 
$ B_{\ell}^{m}(\eta)=-|e|C_{\ell}^{m}\langle r^{\ell}\rangle \alpha_{\ell}h_{\ell}^{m}(\eta)$, 
where the explicit forms of $ C_{\ell}^{m}$ and $ h_{\ell}^{m}(\eta)$ are defined in ref.~\cite{WM2021}. 
The Stevens factors for Tb$^{3+}$ are given as $\alpha_2=-1/99$, $\alpha_4=2/16335$, and $\alpha_6=-1/891891$~\cite{Hutchings}. 
The Dirac-Fock calculation for Tb$^{3+}$ yeilds $\langle r^2\rangle=0.2302$~\AA$^2$, $\langle r^4\rangle=0.1295$~\AA$^4$, and $\langle r^6\rangle=0.1505$~\AA$^6$~\cite{Freeman}. 
Since the Stevens operators $O_{\ell}^{m}(\eta)$ are given in ref.~\cite{WM2021}, the matrix element of $H_{\rm CEF}$ for Tb$^{3+}$ are obtained for the basis set of the eigenstates of the total angular momentum $J=6$ and the $z$ component $|J=6, J_z \rangle$ with $J_z=6, 5, \cdots, -5$, and $-6$. 

\subsection*{Magnetically ordered state in quasicrystal}

In the calculation of (\ref{eq:HI}) applied to the Cd$_{5.7}$Yb-type QC~\cite{Takakura}, we have evaluated the ground-state energy of the model (\ref{eq:HI}) on the ICs located at each vertex of the $\tau^3$-times enlarged icosidodecahedron from that in Fig.~\ref{fig:atoms}D. 
Here, $\tau=(1+\sqrt{5})/2$ is the golden ratio.  
In the icosidodecahedron, there exist 30 vertices at which 30 ICs are located. For the neighboring pairs of ICs, whose number is 60, we set the N.N. interaction $J_1$ and N.N.N. interaction $J_2$ between the magnetic moments at the N.N. and N.N.N. Tb sites respectively. 
In the region shown in Fig.~\ref{fig:QC_mom}A, we have confirmed that the FM arrangement of the magnetic textures on the IC has the lowest energy. Namely, the magnetic state shown in Fig.~\ref{fig:PD}B on the IC located at the origin is surrounded by the same magnetic states on the 30 ICs located at each vertex of the icosidodecahedron (see Fig.~\ref{fig:QC_mom}B) and they are repeatedly arranged outward in the self-similar manner. Hence the FM long-range order is realized in the QC.

\subsection*{Data Availability}

All study data are included in the article and/or SI.

\subsection*{Acknowledgments}

This work was supported by JSPS KAKENHI Grant Numbers JP18K03542 and JP19H00648.


\newpage



\end{document}